\documentstyle [12pt,epsf]{article}

\input{epsf}

\begin{document}

\begin{titlepage}

\begin{center}

{\Large\bf{ Dimensional reduction of the chiral-continous 
Gross-Neveu model } }\\ 

\vspace{.3in}{\large\em 
G.N.J.A\~na\~nos, A.P.C.Malbouisson, M.B.Silva-Neto 
and N.F.Svaiter*}\\
 Centro Brasileiro de Pesquisas Fisicas-CBPF\\
 Rua Dr.Xavier Sigaud 150, Rio de Janeiro, RJ 22290-180 Brazil 

\end{center}

\begin{abstract}

We study the finite-temperature phase transition of the generalized 
Gross-Neveu model with continous chiral symmetry in $2 < d \leq 4$ 
euclidean dimensions. The critical exponents are computed to the leading 
order in the $1/N$ expansion at both zero and finite temperatures. 
A dimensionally reduced theory is obtained after the introduction of thermal 
counterterms necessary to cancel thermal divergences that arise in
the limit of high temperature. Although at zero temperature we have an 
infinitely and continously degenerate vacuum state, we show that at finite 
temperature this degeneracy is discrete and, depending on the values of the 
bare parameters, we may have either total or partial restoration of symmetry.
Finally we determine the universality class of the reduced
theory by a simple analysis of the infrared structure of thermodynamic
quantities computed using the reduced action as starting point.
 
\end{abstract}

Pacs numbers: 11.10.Ef, 11.10.Gh 

\end{titlepage}

\newpage\baselineskip .37in


\section{Introduction}

One of the differences between field theories and mechanical systems
is that field theories have an infinite number of degrees of freedom,
which makes possible spontaneous symmetry breaking. More generally, we
know, by now, that also dynamical
symmetry breaking plays an important role in modern physics. In 
condensed matter physics, for example, such a mechanism is used to describe 
superconductivity, by the condensate of Cooper pairs, while 
for particle physics it is the main source 
of hadron masses and governs the low energy hadron dynamics.

Connected with dynamical symmetry breaking comes the problem
of restoration of broken symmetries for some sufficiently high temperature.
As it is well known, a phase transition occurs when there is a 
singularity in the free energy or one of its derivatives.
The fact that almost all phenomena studied in theories near the 
transition point exhibit scaling, that is, a power-law behavior 
between two mesurable quantities, leads naturally to the classification
of different conformal field theories in universality classes, 
which are itself determined by a set of numbers usually called
critical indices.

A large number of conformal field theories is known in 
two dimensions and none is firmly established in four. The dimensionality
$(d=3)$ is between these extremes and contains two well established 
families of conformal field theories. The first family contains the
usual Ising, XY and Heisenberg critical $\sigma$ models 
and is characterized by a particular set of exponents 
called scalar critical exponents.
The second family contains various critical four fermion models 
\cite{Gat-Kovner-Rosenstein} and the set of exponents that
characterizes it is called chiral critical exponents.
The critical properties of the great majority of phase transitions 
in three euclidean dimensions (magnetic systems, superconductors etc.) 
are quite accurately described by the first family of conformal field 
theories while the finite-temperature phase transitions in certain $(3+1)$ 
dimensional quantum field theories are argued to belong to these same 
universality classes \cite{Svetitsky-Yaffe}. 

When studying the finite-temperature chiral 
restoration in QCD one is usually gided by the concepts of dimensional
reduction and universality. From the dimensional reduction point of view,
a hot field theory can be regarded as a static field theory at zero
temperature in a lower dimension \cite{Jourjine}.
For example, four dimensional QCD with $N_{f}$ light quarks 
near the transition can be described by the three dimensional linear $\sigma$ model with the same global symmetry \cite{Wilczek}. 
On the other hand, as the classification in universality classes is done
by the computation of critical indices and
those critical indices are infrared (IR) sensitive, we must understand
the role of each degree of freedom in this context. This can be easily
done with the use of effective field theory methods in which the 
integration over different energy scales gives a theory for 
the degrees of freedom which are the real responsible for infrared 
divergencies \cite{Braaten}. Dimensional reduction is based in the
zero temperature Appelquist-Carazone decoupling theorem
\cite{Appelquist-Carazone}, where a low energy theory is constructed
by the integration over heavy fields in the functional integral.

According to standard dimensional reduction arguments, 
the fermions themselves, even if they
are massless at zero temperature, do not influence the nature of phase
transition at finite-temperature. It is rather their bosonic composites, 
the Goldstone bosons, which are of importance. This follows directly from 
the universality of second order phase transitions \cite{Amit}, in 
which the commonly held assumption
is that all the possible universality classes (or equivalently, 
conformal field theories) are variations of the $\sigma$ model and one
need only match the correct symmetry-breaking pattern. As a consequence,
the chiral transition of four dimensional QCD, with $N_{f}=2$ flavors,
should lie in the same universality class as a three dimensional $O(4)$
magnet. Similarly, other models, e.g. four-fermion theories in $d$
dimensions such as the Gross-Neveu model (GN model) \cite{Gross-Neveu} 
with discrete symmetries or the Nambu-Jona-Lasinio model (NJL model) 
\cite{Nambu-Jona-Lasinio} with continous chiral symmetry, 
are expected to be in the same universality class of a $(d-1)$ dimensional
Ising or Heisenberg magnet, respectively.

Recently, however, it was pointed out that there exist different $(d=3)$
conformal field theories with the same symmetry-breaking pattern 
\cite{Gat-Kovner-Rosenstein}.
They are exactly the four-fermion interaction models of the 
Nambu-Jona-Lasinio type.
Physically, this corresponds to the fact that on the chirally symmetric
side of the phase transition there are $N$ massless fermions whose effect
is felt even in the IR fixed point, just like the Goldstone bosons.
The presence of more than one universality class in $(d=3)$ makes the
procedure of dimensionally reducing a quantum field theory ambigous
and it is now uncertain to which
conformal field theory the $(d+1)$ finite-temperature quantum
field theory will reduce. 
The argument if favor of the bosonic universality class goes as follows.
At finite-temperature, the $(d=4)$ fermion reduces to a collection of 
$(d=3)$ massive fermions and there is no zero mode for which the 
Matsubara frequency vanishes. Nevertheless,
even if a single massive field does not influence the phase transition, the cumulative effects of an infinite number of such fields may 
have an appreciable impact. In order to see whether or not this happens,
all harmonics should be summed and their cumulative effects studied.

It is the purpose of this paper to discuss the assumptions underlying this
analysis and to determine to which universality class the simplest generalization of the Gross-Neveu model with continous chiral symmetry,
belongs. Besides being an interesting theoretical model, it is also
believed that, when properly extended to incorporate continous chiral
symmetry, four-fermion models are more realistic as effective theories
of QCD than the linear $\sigma$ model, especially at scales where
quark structure is important.

The paper is organized as follows. In section II, we present the model. In 
section III we compute the critical exponents at zero temperature while, in 
section IV, we obtain, after dimensional reduction, the new critical exponents 
that determines the universality class of the reduced theory. In section V, we 
analyse the vacuum structure at finite-temperature. Conclusions are given in 
section VII. In the appendix, we compute the thermal renormalization group 
functions that controlls thedependence of the thermal counterterms on the 
temperature. In this paper we use $h\!\!\!\slash=c=1$.


\section{The chiral-continous GN model}

We are interested in studying the behavior of a multiplet of
N fermions coupled with a pair of composite self-interacting pseudo-scalar 
and scalar fields, $\sigma(x)$ and $\pi(x)$ respectively, 
in such a way that the euclidean functional action reads
\begin{equation}
S_{E}=\int d^{d}{\bf x}
\left\{
-\bar{\psi}
\left( \partial\!\!\!\slash+ 
       g_{B}\left(\sigma+i\gamma_{S}\pi 
        \right)
\right)
\psi+
\frac{1}{2}(\partial_{\mu}\sigma)^{2}+
\frac{1}{2}(\partial_{\mu}\pi)^{2}-
\frac{1}{2}m_{B}^{2} \left( {\sigma}^{2}+{\pi}^{2} \right) +
\frac{\lambda_{B}}{4} \left( {\sigma}^{2}+{\pi}^{2} \right)^{2}
\right\}.
\label{initial-action}
\end{equation}
This model has continous chiral symmetry
\begin{equation}
\left\{ \begin{array}{ll}
        \bar{\psi} \rightarrow \bar{\psi}e^{i\alpha \gamma_{s}} \\
        \psi \rightarrow e^{i\alpha \gamma_{s}} \psi,
        \end{array}
\right.,
\end{equation}
and $\sigma \rightarrow -\sigma$, $\pi \rightarrow \pi$ with $\alpha$
being a constant and we use for the $\gamma_{s}$ the following representation 
$\gamma_{S}=
\left(
\begin{array}{ll}
1 &  0 \\
0 & -1
\end{array}
\right)
$.

In order to have an uniquely defined vacuum (see fig. 1), we must choose a 
prefered direction in the space of components, say  $\left< \sigma \right>=v$ 
and $\left< \pi \right> = 0$, and shift our fields as
\begin{equation}
\left\{ \begin{array}{ll}
        \sigma(x) \rightarrow v + \sigma(x)  \\
        \pi(x)    \rightarrow \pi(x)
        \end{array}
\right.,
\end{equation}
where $v=\frac{\sqrt{m_{B}}}{\lambda_{B}}$. In this new picture the euclidean 
action becomes
\begin{eqnarray}
S_{E} & = & \int d^{d}{\bf x}
\left\{ 
-\frac{1}{2}m_{B}^{2}v^{2}+
\frac{\lambda_{B}}{4}v^{4}-
\bar{\psi}
\left( \partial\!\!\!\slash+ 
       g_{B}\left(v+\sigma+i\gamma_{S}\pi 
        \right)
\right)
\psi+
\frac{1}{2}(\partial_{\mu}\sigma)^{2}+
\frac{1}{2}(\partial_{\mu}\pi)^{2} \right. \\
& - &
\left.
\frac{1}{2}m_{\sigma}^{2}\sigma^{2}-
\frac{1}{2}m_{\pi}^{2}\pi^{2}-
{\lambda_{B}}v{\sigma^{3}}-
{\lambda_{B}}v{\sigma}{\pi^{2}}+
\frac{\lambda_{B}}{4} \left( {\sigma}^{2}+{\pi}^{2} \right)^{2}
\right\},
\label{shifted-action}
\end{eqnarray}
with $m_{\sigma}=m_{B}^{2}-3\lambda_{B}v^{2}$ and
$m_{\pi}=m_{B}^{2}-\lambda_{B}v^{2}$. If we use the definition of $v$ we can
see that $m_{\pi}=0$, as it should be since the $\pi$ field is a Goldstone
boson. Although new interaction terms have arised, it is easy to prove that
this renormalizable model needs only usual wave function, mass and coupling
constant renormalization to render the theory finite.

%
\begin{figure}[th]
  
\centerline{\epsfysize=3.5in\epsffile{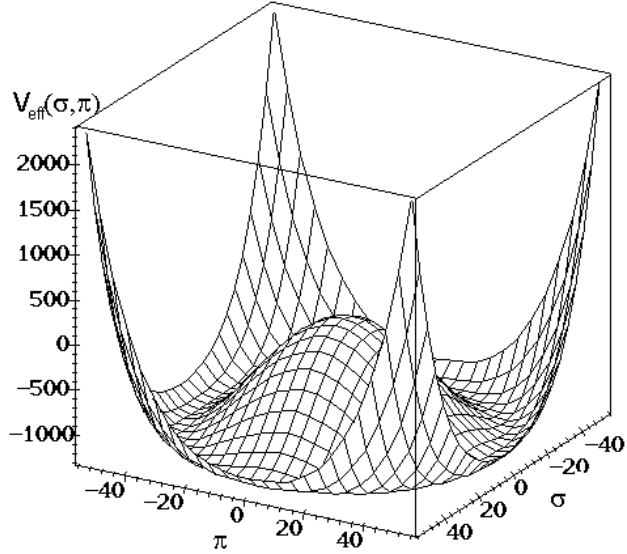}} 
\caption[region]
{\small\sf{Mexican hat shape for the zero temperature effective potential
in the tree level. The vacuum is infinitely degenerate and one 
of the degrees of freedom will become a Goldstone boson after
the definition of an unique ground-state.}}
\end{figure}

The finite-temperature version of this model is obtained, in the Matsubara
formalism, by the compactification of the imaginary time dimension together
with the imposition of (anti)-periodic boundary conditions to 
(fermionic)-bosonic fields \cite{Jourjine} as
\begin{equation}
\bar{\psi}({\bf x},\tau)=
\frac{1}{\sqrt{\beta}}
\sum_{n=-\infty}^{\infty}\bar{\psi}_{n}({\bf x})e^{i\omega_{n}\tau},
\end{equation}
with $\omega_{n}=\frac{2\pi}{\beta}(n+\frac{1}{2})$ and
\begin{equation}
\sigma({\bf x},\tau)=\frac{1}{\sqrt{\beta}}
\sum_{n=-\infty}^{\infty}\sigma_{n}({\bf x})e^{i\omega_{n}\tau},
\end{equation}
with $\omega_{n}=\frac{2\pi n}{\beta}$. 

Since global aspects of the euclidean manifold do not affect local properties 
of a quantum field theory we will, again, need only those previously 
mentioned renormalization constants to render the theory finite. 


\section{Critical exponents at zero temperature}

The chiral-continous Gross-Neveu model, in contrast to the discrete 
Gross-Neveu model, is renormalizable
in four dimensions and it can be described, beyound the tree level,
by all the techniques developed for the $(\Phi^{2})^{2}$ theory: 
renormalization group equations and large N expansion, for example.

To investigate the vacuum structure of the theory, we must consider the 
leading order density $V_{eff}$ evaluated for constant fields 
$\left< \sigma \right>=v$ and $\left< \pi \right>=0$
\cite{Rosenstein-report}. 
This gives the gap equation (a cut-off $\Lambda$ is implied)
\begin{equation}
\frac{\partial V_{eff}}{\partial \sigma}=0 \Longrightarrow
-m_{B}^{2}+\lambda_{B}v^{2}=
2g_{B}^{2}\int^{\Lambda}\frac{d^{d}q}{(2\pi)^{d}}
\frac{1}{q^{2}+g_{B}^{2}v^{2}}.
\end{equation}
We define the critical coupling $g_{c}$ as
\begin{equation}
-m_{B}^{2}=2g_{c}^{2}\int^{\Lambda}\frac{d^{d}q}{(2\pi)^{d}}
\frac{1}{q^{2}},
\end{equation}
so that the gap equation takes the form
\begin{equation}
\frac{1}{2}
\left(
m_{B}^{2}t-\frac{\lambda_{B} v^{2}}{t+1}
\right)
=
\int^{\Lambda}\frac{d^{d}q}{(2\pi)^{d}}
\frac{g_{B}^{2}v^{2}}{q^{2}(q^{2}+g_{B}^{2}v^{2})},
\label{gap-equation}
\end{equation}
where $t=\left(\frac{g_{B}^{2}}{g_{c}^{2}}-1 \right)$ is the reduced coupling.

This form for the gap equation is particularly well suited for extracting 
critical indices since the problem
reduces to counting the infrared divergences on the right hand side
\cite{Zinn-Justin}. As previously said, the
critical indices are the responsible for the power-law behavior 
between two mesurable quantities as we approach the critical coupling. 
In this sense
we will compute the critical exponents defined by 
$\left. \left< \bar{\psi}\psi \right> \right|_{v \rightarrow 0}
\sim t^{\beta}$, 
$\left. \left< \bar{\psi}\psi \right> \right|_{t \rightarrow 0}
\sim v^{1/\delta}$,
$\left. \left< \Sigma \right> \right|_{v \rightarrow 0}
\sim t^{\nu}$ etc, 
and show that the introduction of a self-interaction
for the bosonic fields does not change the values of the exponents
obtained for the NJL model. 

First, we can see that the critical indices for the NJL model are 
recovered by simply putting $\lambda_{B}=0$ in the gap equation 
(\ref{gap-equation}). Indeed, in this case the infrared behavior of 
the integral with $v$ goes as
\begin{equation}
\int_{v}^{\Lambda}\frac{d^{d}q}{(2\pi)^{d}}
\frac{g_{B}^{2}v^{2}}{q^{2}(q^{2}+g_{B}^{2}v^{2})}
\sim
g_{B}^{2}v^{d-2}-g_{B}^{4}v^{d-4}+g_{B}^{6}v^{d-6}-...,
\label{integral-behavior}
\end{equation}
and the critical indices are simply $\beta=\nu=1/(d-2)$ and $\delta=d-1$, 
since, to the order we are concerned, the self-energy and the
two point function coincide.

Now, let us show that even with $\lambda \neq 0$ we still have the same 
critical indices. For $d<4$ this is rather obvious since the singularity 
of the integral in (\ref{gap-equation}) allow us to neglect the $v^{2}$ 
term in the left hand side. However, if we are in four dimensions, 
we can use the integral behavior (\ref{integral-behavior}) to obtain
\begin{equation}
m_{B}^{2}t-\frac{\lambda_{B} v^{2}}{t+1} \sim 2g_{B}^{2}v^{2},
\end{equation}
and compute the $\beta$ exponent from the behavior of
\begin{equation}
v^{2} \sim \frac{m_{B}^{2}t}{\frac{\lambda_{B}}{t+1}+2g_{B}^{2}}
\end{equation}
with $t$. This gives the mean field value $\beta=\frac{1}{2}$, 
a characteristic of conformal field theories in dimensions greater 
or equal to 4. 

We conclude that the introduction of self-interaction for the composite 
bosons, has not changed the structure of the IR singularities. 
Also, we can see that, in $(d=4)$ all critical exponents are mean 
field and there are no non-gaussian fixed points while, in three 
dimensions the value $\beta=1$ tell us this is a chiral conformal 
field theory of the Nambu-Jona-Lasinio type.


\section{Dimensional reduction and universality}

We will now consider the problem of computing critical indices when
a non zero temperature is introduced. We could do this in a similar way
to what we have done in the zero temperature case but we will, instead,
compute the fermionic determinant using a modified minimal subtraction
scheme at zero momentum and temperature $1/\beta$ because this is a more
suitable way of obtaining the functional form of the dimensionally 
reduced theory.

As we have pointed out in the introduction, the reduced theory is an 
effective theory for the zero modes of the original fields
\cite{Jourjine} (and see also \cite{Landsman}).
It can be explicitly obtained from the 
initial partition funcion by integrating out the $\omega_{n} \neq 0$
Matsubara frequencies of the bosonic fields and, as we do not have
a zero mode in the case of fermions, the complete integration
over the $\psi$ and $\bar{\psi}$ fields. The resulting QFT will be
described by an euclidean partition function of the type
\begin{equation}
Z=\int {\cal D}{\sigma_{0}}{\cal D}{\pi_{0}} \,
exp{ \left\{
-\int d^{d-1}{\bf x}
\left(
{\cal L}_{eff}(\sigma_{0},\pi_{0})+\delta{\cal L}
\right)
     \right\} }, 
\end{equation}
where $\delta{\cal L}$ includes all other local terms 
not present in the original lagrangian density but which are 
consistent with the symmetries and, in the lagrangian density 
${\cal L}_{eff}(\sigma_{0},\pi_{0})$, the parameters will be, in
general, functions of the cut-off, the temperature and the bare
parameters.

Since we are interested in the computation of the finite-temperature 
effective action at the leading order in the $1/N$ expansion it is
convenient to define
\begin{equation}
\left\{ \begin{array}{ll}
        \tilde{g}=g_{B} \sqrt{N} \\
        \tilde{m}^{2}=m_{B}^{2}N \\
        \tilde{\lambda}=\lambda_{B} N.
        \end{array}
\right., 
\end{equation}
and
\begin{equation}
\left\{ \begin{array}{ll}
        \bar{m}_{1}=\tilde{g}_{B}\left( \sigma_{0}+i\pi_{0} \right) \\
        \bar{m}_{2}=\tilde{g}_{B}\left( \sigma_{0}-i\pi_{0} \right),
        \end{array}
\right.
\end{equation}
being the eigenvalues of 
$\bar{M}\equiv g_{B}({\bf I}\sigma_{0}+i\gamma_{S}\pi_{0})$,
where ${\bf I}$ is the unit matrix.

At finite-temperature, the integration over the imaginary
time becomes a sum over Matsubara frequencies and if we remember that the 
fields $\bar{\psi}$ and $\psi$ are anti-periodic in the time component, we obtain, after integrating out the fermionic degrees of freedom, the expression
for the fermionic determinant 
\begin{equation}
trln{\left(\partial\!\!\!\slash+\bar{M}\right)}=
\frac{2}{\beta} \sum_{n=-\infty}^{\infty}
\int \frac{d^{d-1}{\bf k}}{(2\pi)^{d-1}}
\sum_{i=1}^{2}\ln{\left( {\bf k}^{2}+\omega_{n}^{2}+\bar{m}_{i}^{2}\right)},
\label{tr-ln-1}
\end{equation}
where, as usual, $\omega_{n}=\frac{2\pi}{\beta}(n+\frac{1}{2})$ for fermions,
or yet
\begin{equation}
trln{\left(\partial\!\!\!\slash+\bar{M}\right)}=
\frac{2}{\beta} \sum_{n=-\infty}^{\infty}
\int \frac{d^{d-1}{\bf k}}{(2\pi)^{d-1}}
\left\{
2\ln{\left( {\bf k}^{2}+\omega_{n}^{2} \right)}+
ln{\left( 1+\frac{\bar{m}_{1}^{2}}{{\bf k}^{2}+\omega_{n}^{2}}\right)}+
ln{\left( 1+\frac{\bar{m}_{2}^{2}}{{\bf k}^{2}+\omega_{n}^{2}}\right)}
\right\}.
\label{tr-ln-2}
\end{equation}
The first logarithim in the above expression is naturally absorved as an
overall normalization factor. The other two can be written, after 
integration over ${\bf k}$ and summation over $n$, as a series
\begin{equation}
trln{\left(\partial\!\!\!\slash+\bar{M}\right)}=
2 \sum_{s=1}^{\infty} \frac{(-)^{s}{\tilde{g}}^{2s}}{s}
\left[ \left( \sigma_{0}+i\pi_{0} \right)^{2s}+
       \left( \sigma_{0}-i\pi_{0} \right)^{2s}
\right]
\frac{(2\pi)^{d-1-2s}}{(4\pi)^{\frac{d-1}{2}}}
\frac{\Gamma\left( s-\frac{d-1}{2}\right)}{\Gamma(s)}
\frac{\zeta \left( 2s-(d-1),\frac{1}{2}\right)}{\beta^{d-2s}},
\label{tr-ln-3}
\end{equation}
where $\zeta$ is the analytic extension of the modified 
Epstein-Hurwitz zeta function \cite{Elisalde-Romeo}.

In the limit of high temperature ($\beta \rightarrow 0$) the only 
contributions to the effective action comming from expression
(\ref{tr-ln-3}) are those given by $s=1$ and $s=2$. Those contributions
are divergent so that thermal couterterms will be needed to make the 
reduced theory finite in this limit \cite{Landsman}. 
In this context, we join
the above divergences with those comming from the integration over 
non-static bosonic modes \cite{Malbouisson-Silva-Neto-Svaiter} and
introduce the following counterterms (for $d=4-\varepsilon$) 
\begin{equation}
\delta m^{2}=\frac{1}{24}\frac{1}{\beta^{2}}
\left( 4\tilde{\lambda}{\bf I}-\tilde{g}^{2}\gamma_{S} \right)+
\frac{\tilde{\lambda}\tilde{m}^{2}{\bf I}}{4\pi^{2}}
\left( 
\frac{1}{\varepsilon}+1-\gamma-ln{\frac{2\beta\mu}{\sqrt{\pi}}}
\right),
\label{mass-counterterm}
\end{equation}
and
\begin{equation}
\delta \lambda=
\left[
\frac{10}{6}\frac{\lambda^{2}}{64\pi^{2}}{\bf I_{4}}+
\frac{{\tilde{g}^{4}}}{16\pi^{2}}
\left(
\begin{array}{ll}
       1 & -3 \\
      -3 &  1
\end{array}
\right)
\right]
\left(
-\frac{1}{\varepsilon}+\frac{\gamma}{2}+ln{\frac{2\beta\mu}{\sqrt{\pi}}}
\right),
\label{coupling-counterterm}
\end{equation}
where, as usual, $\gamma$ is the Euler number, $\mu$ is a 
mass parameter introduced by dimensional regularization to give Green 
functions their proper dimensions and we have defined, for simplicity
\begin{equation}
{\bf I_{4}}
=
\left(
\begin{array}{ll}
       1 & 1 \\
       1 & 1
\end{array}
\right).
\end{equation}

Since as $\beta \rightarrow 0$ we are in the disordered phase 
($v=0$), we define
\begin{equation}
\phi=
\left(
\begin{array}{c}
\sigma_{0} \\
\pi_{0}
\end{array}
\right),
\end{equation}
and
\begin{equation}
\Phi=
\left(
\begin{array}{ll}
\sigma_{0} & 0 \\
0          & \pi_{0}
\end{array}
\right)
,
\end{equation}
so that the renormalized action for the $(d=4)$ becomes
\begin{equation}
S_{eff}({\sigma_{0}},{\pi_{0}})=\int d^{3}{\bf x}
\left\{
\frac{1}{2}\left( \partial_{i}\phi^{a} \right)^{2}+
\frac{1}{2}\phi^{a}m_{ab}^{2}(\beta)\phi^{b}+
\frac{1}{4}\phi^{a}
           \Phi_{a\alpha}
           \frac{\lambda_{\alpha \alpha^{'}}(\beta)}{\beta}
           \Phi_{\alpha^{'}b}
           \phi^{b}+
\frac{28}{9}
\frac{\tilde{\lambda}^{3}\zeta(3)}{2^{9}\pi^{4}}(\phi^{a}\phi^{a})^{3}
\right\},
\label{reduced-action}
\end{equation}
where, as usual,
\begin{equation}
\left\{
\begin{array}{ll}
m^{2}(\beta)=-\tilde{m}^{2}{\bf I}+\delta m^{2} \\
\lambda(\beta)=
\tilde{\lambda}{\bf I_{4}}+
\delta \lambda.
\end{array} 
\right.
\label{renormalized-parameters}
\end{equation}

Now we are ready to start the computation of the critical indices for
the reduced theory. In this sense we will, again, set 
$\left<\pi\right>=0$ and expand our field $\sigma$ around a constant 
configuration $v$ to obtain the thermal effective potential 
\begin{equation}
V_{eff}=
\frac{1}{2}m_{10}^{2}(\beta)v^{2}+
\frac{1}{4}\frac{\lambda_{10}(\beta)}{\beta}v^{4}+
\frac{28}{9}
\frac{\tilde{\lambda}^{3}(\beta)\zeta(3)}{2^{9}\pi^{4}}v^{6}.
\label{effective-potential}
\end{equation}

The first critical exponent one usualy computes is the $\beta$ exponent 
which gives the power-law behavior of the order parameter with the 
temperature. The critical temperature $\beta_{c}$ is
itself determined by the requirement of a vanishing thermal mass when 
$\beta \rightarrow \beta_{c}$. In this sense the condition 
$m_{10}(\beta_{c})=0$ gives
\begin{equation}
\beta_{c}^{2}=
\frac{1}{24}
\frac{\left( 4\tilde{\lambda}-{\tilde{g}}^{2}\right)}{\tilde{m}^{2}}.
\end{equation}
In order to have a critical temperature we will, for the moment, consider
the case $4\tilde{\lambda}>\tilde{g}^{2}$. Now, we can rewrite 
$m_{10}^{2}(\beta)$ as
\begin{equation}
m_{10}^{2}(\beta)=
\frac{1}{24}\left( 4\tilde{\lambda}-\tilde{g}^{2} \right)
\left( \frac{1}{\beta^{2}}-\frac{1}{\beta_{c}^{2}} \right).
\end{equation}
As $\beta \rightarrow \beta_{c}$ we see that, by defining the reduced
temperature $\theta$ as
\begin{equation}
\beta=\frac{\beta_{c}}{1+\theta \beta_{c}},
\end{equation}
we obtain
\begin{equation}
m_{10}^{2}(\beta)=
\frac{1}{24}\left( 4\tilde{\lambda}-\tilde{g}^{2} \right)
\frac{2\theta}{\beta_{c}}+O(\theta^{2}) \sim a+b\theta,
\end{equation}
that is, $m_{10}^{2}(\beta)$ is linear in $\theta$, or equivalently,
we have for the first critical exponent the value $\beta=\frac{1}{2}$. 

The second critical exponent, the $\delta$
critial exponent, is obtained from the inhomogenious gap equation,
when we consider an external source $J$ for the $\sigma$ field. In this case, 
the gap equation becomes
\begin{equation}
m_{10}^{2}(\beta)v+\frac{\lambda_{10}(\beta)}{\beta}v^{3}+
\frac{28}{3}\frac{\tilde{\lambda}^{3}\zeta(3)}{2^{8}\pi^{4}}v^{5}=J,
\end{equation}
and, as we approach criticality $(m_{10}(\beta_{c}) \rightarrow 0)$, 
the gap equation gives $\delta=3$, since $v^{3}$ dominates the $v^{5}$ 
term. Finally, as we are in the leading order in the $1/N$ expansion, 
once again the two point function and the self-energy are the same. 
This tells that the $\nu$ critical exponent is also equal to $\beta$
assuming the value $\frac{1}{2}$.

After all, to which universality class does the reduced theory belong?
As we have seen, all critical exponents are mean field
in the leading order of the $1/N$ expansion. Actually, one would get the 
same critical exponents in this approximation in $d=3$
\cite{Rosenstein-Warr-Park}, although they would obviously not
satisfy hyperscaling relations in this dimension. Notice also that they
are almost as far from the scalar critical exponents $\beta \approx 0.3$
in three dimensions \cite{Amit}, as from the chiral critical exponents 
computed in the previous section $\beta=1$ \cite{Gat-Vasilev}. 

We would be tempted to conclude that dimensional reduction really spoiled 
the critical behavior of thermodinamic quantities 
\cite{Kocic-Kogut}, reforcing the thesis that this procedure is
ambigous. However, this sentence must be used with care.
The critical exponents we have just computed are related to an
effective field theory obtained from the integration over all fermions
and all nonstatic bosons. Since the critical indices come from the IR behavior 
of mesurable quantities near the transition point and the integration we
have done gives IR finite results, it is reasonable to expect that the above 
degrees of freedom will not improve mean field values for the exponents. In 
order to recover non trivial values for the critical indices we must consider 
the effects of the zero modes $\sigma_{0}$ and $\pi_{0}$ in the computation
of thermodynamic quantities, because they are the only responsible for
IR divergences in the finite-temperature version of this model. 

Following the same steps of Kogut in \cite{Kocic-Kogut} we will consider
the critical behavior of the susceptibility $\chi$ for a $\sigma$ model
with action given by eq. (\ref{reduced-action}). Defining the critical
curvature $\mu^{2}_{c}$ as the point where the susceptibility diverges
\begin{equation}
\mu^{2}_{c}=\frac{\lambda_{10}(\beta)}{\beta} 
\int^{\Lambda}\frac{d^{d-1}q}{(2\pi)^{d-1}}
\frac{1}{q^{2}},
\end{equation}
the expression for the inverse susceptibility can be recast into
\begin{equation}
\chi^{-1}
\left(
1+\frac{\lambda_{10}(\beta)}{\beta}\int^{\Lambda}
\frac{d^{d-1}q}{(2\pi)^{d-1}}
\frac{1}{q^{2}(q^{2}+\chi^{-1})}
\right)
=
\mu^{2}-\mu^{2}_{c}.
\label{susceptibility}
\end{equation}
Again, the extraction of the critical index $\gamma$ reduces to counting
the powers of the infrared singularities in the left hand side of the
previous equation. In three dimensions $(d-1=3)$ the second term in eq.
(\ref{susceptibility}) dominates the scaling region, giving the 
zero temperature susceptibility exponent $\gamma=2/(d-3)$. 
It is easy to obtain the other critical exponents
and they show the same type of behavior as $\gamma$. The reduced theory
has scalar critical indices and due to the dimensionality of the order
parameter we conclude that it belongs to the universality class 
of the Heisenberg magnet, as reasonably expected.


\section{Vacuum structure at finite-temperature}

In the previous section, we defined the vacuum state in the $\sigma_{0}$
direction, breaking spontaneously a symmetry, and computed the critical 
indices from the effective potential (\ref{effective-potential}), obtained 
after dimensional reduction. Now comes the question. Is this really
a choice? As it will become clear soon, this is definitely not a choice.

If we still had the shape of a "mexican hat" for the effective potential
(see fig. 1), we would simply have to choose a prefered direction to be
the minimum configuration defining the vacuum of the theory. However,
differently from the zero temperature situation, this is no longer the case. 
As we can see from expression (\ref{mass-counterterm}), the integration over 
fermi fields give different thermal contributions for the masses of 
$\sigma_{0}$ and $\pi_{0}$ and if we define the critical temperature as
the temperature in which we have a vanishing mass, we will conclude that,
since we have two different masses, we will have two critical temperatures
$\beta_{\sigma}^{-1}$ and $\beta_{\pi}^{-1}$ such that

%
\begin{figure}[hp]
\centerline{\epsfysize=4in\epsffile{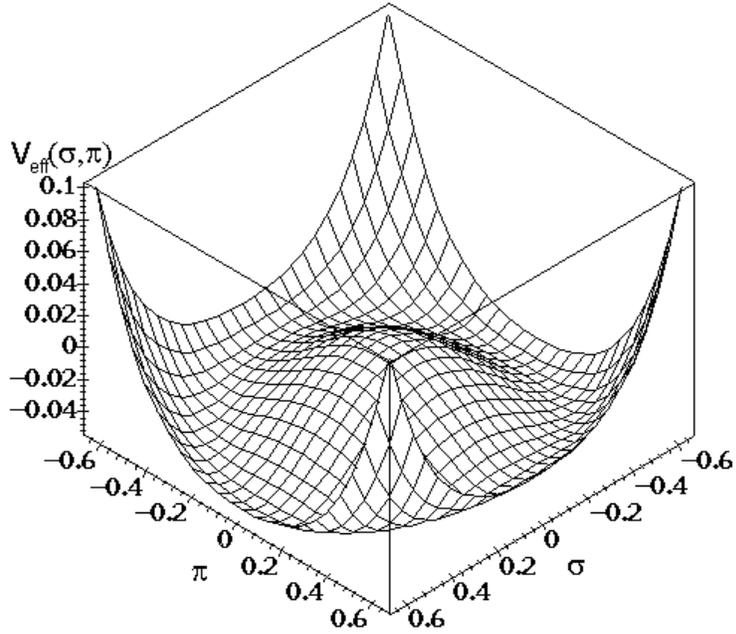}} 
\caption[region]
{\small\sf{Vacuum structure for $\beta>\beta_{\pi}$. We see that the ground state
is twice degenerate in the $\sigma_{0}$ direction while a false vacuum 
(or metastable vacuum) can be identified in the $\pi_{0}$ direction.
After the identification of the true vacuum state we will no longer
have a Goldstone boson.}}
\end{figure}
\begin{equation}
\beta_{\sigma}^{2}=
\frac{1}{24}
\frac{\left( 4\tilde{\lambda}-{\tilde{g}}^{2}\right)}{\tilde{m}^{2}}
\end{equation}
and
\begin{equation}
\beta_{\pi}^{2}=
\frac{1}{24}
\frac{\left( 4\tilde{\lambda}+{\tilde{g}}^{2}\right)}{\tilde{m}^{2}}.
\end{equation}

Now, let us show that associated to each critical temperature 
there is a second order phase transition and a symmetry restoring 
phase. Indeed, if we go from zero temperature $(\beta = \infty)$ to a 
temperature close to $\beta_{\pi}^{-1}$ the effective potential takes 
the form of (fig. 2). There is a discrete degeneracy of the vacuum 
showing that in the previous section we made a correct choice of the 
ground-state.

If we continue increasing the temperature in such a way that 
$\beta_{\sigma}<\beta<\beta_{\pi}$ we note (see fig. 3) that
the discrete symmetry in the $\pi_{0}$ direction is restored but
we continue with a broken phase in the $\sigma_{0}$ direction.
Complete restoration of symmetry (fig. 4) will only be achieved 
when $\beta<\beta_{\sigma}$, and this is exactly the picture after 
dimensional reduction since we have already taken the limit 
$\beta \rightarrow 0$.

%
%
\begin{figure}[th]
\centerline{\epsfysize=4in\epsffile{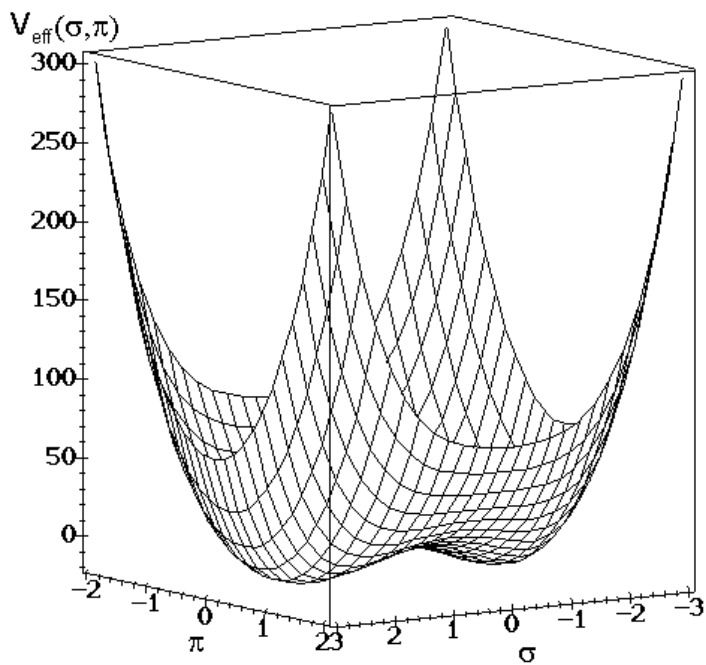}} 
\caption[region]
{\small\sf{Partial restoration of symmetry for $\beta_{\sigma}<\beta<\beta_{\pi}$. 
The $\pi_{0}$ degree of freedom is now in the disordered phase while 
we still have discrete breakdown of symmetry in the $\sigma_{0}$
direction.}}
\end{figure}

All the above discussion is valid only if $4\tilde{\lambda}>\tilde{g}^{2}$.
What happens when $4\tilde{\lambda}<\tilde{g}^{2}$? In this case
we will have only the $\beta_{\pi}$ critical temperature. This is to say,
the discrete symmetry on the $\pi_{0}$ direction will be restored while,
in the $\sigma_{0}$ direction, the minimum will become deeper and deeper
as we increase the temperature. We will not have a complete restoration of 
symmetry and the effective potential will have the shape of (fig. 3).
%
%
\begin{figure}[th]
\centerline{\epsfysize=4in\epsffile{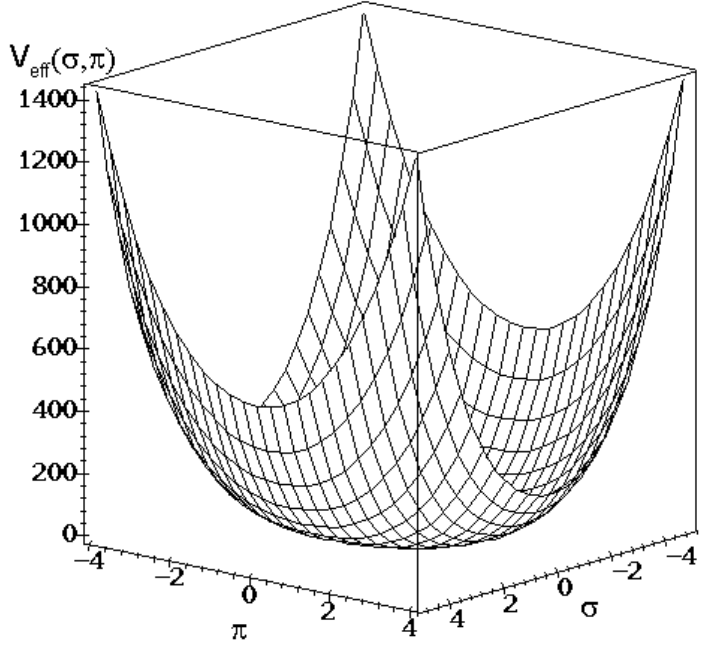}} 
\caption[region]
{\small\sf{Complete disorder for $\beta<\beta_{\sigma}$. We see that, due to the
difference in the thermal masses of the static fields $\sigma_{0}$ and
$\pi_{0}$ this surface cannot be obtained by the revolution of a 
generating curve.}}
\end{figure}
%


\section{Conclusions}

Dimensional reduction of the chiral-continous Gross-Neveu model was
performed giving an effective theory that belongs to the same universality
class of the Heisenberg magnet. In spite of the integration over fermions 
and nonstatic bosons modes have given mean field results for the critical indices, we were able to explicitely determine the universality
class of the reduced theory by simply taking into account the effects of
the only relevant degrees of freedom in the computation of thermodynamic
quantities; the static Matsubara modes.

The presence of a coupling $i\gamma_{S}$ in the
original action is the responsible for the difference on the thermal
contributions for the masses of the pseudo-scalar and scalar fields 
$\sigma_{0}$ and $\pi_{0}$. This difference has important implications. 
First, we saw that if $4\tilde{\lambda}>\tilde{g}^{2}$
we have restoration of symmetry for some critical temperature, first
in the $\pi_{0}$ direction and after in the $\sigma_{0}$ direction, 
while, if $4\tilde{\lambda}<\tilde{g}^{2}$ we have only partial restoration
of symmetry. The introduction of temperature in a theory with a continous 
chiral symmetry has changed the role played by the $\pi_{0}$ field from the
zero temperature case, where it was a Goldstone boson, to the 
finite-temperature case, where this is no longer true.


\section{Acknowledgment}

We are greatfull to C. de Callan, S. A. Dias and R. O. Ramos for 
interesting discussions. This paper was suported by Conselho Nacional de 
Desenvolvimento Cient\'{\i}fico e Tecnol\'ogico do Brasil (CNPq) and 
Funda\c c\~ao Coordenadoria de Aperfei{\c c}oamento de Pessoal de 
N\'{\i}vel Superior (CAPES).


\section*{Appendix A - Thermal RG flow}

We compute, for completeness, the thermal renormalization group 
functions which controlls the dependence of the counterterms
on the temperature. As it is well known \cite{Landsman}, the transition 
from a four-dimensional theory to an effective three-dimensional one is 
renormalization point dependent. This is not surprising since the 
Appelquist-Carazone decoupling theorem \cite{Appelquist-Carazone}
only holds if a particular class of renormalization prescriptions
is adopted, being optimal in BPHZ subtractions at zero momenta and
temperature $\beta$. Neverthless, we can write down the thermal
renormalization group equation which controls the dependence with
the temperature on the correlation functions and compute the 
running quantities.

The independence of bare correlation functions on the choice of the 
renormalization point, is expressed by the conditions
\begin{eqnarray}
\mu\frac{d}{d\mu}\Gamma^{(N)}=0, \\
T\frac{d}{dT}\Gamma^{(N)}=0,
\end{eqnarray}
and, for simplicity, we will work, in this section, with the temperature 
$T$ instead of $\beta \equiv 1/T$.

Since we are only interested on the computation of thermal 
renormalization group functions, we will consider solely
\begin{equation}
\left(
         T_{0}\frac{\partial}{\partial T_{0}}+
         \beta_{T}\frac{\partial}{\partial\lambda}+
         \gamma_{T}m^{2}\frac{\partial}{\partial m^{2}}
\right)
\Gamma^{(N)}=0
\label{rg-equation}
\end{equation}
where
\begin{equation}
\beta_{T}=T_{0}\frac{\partial\lambda(T)}{\partial T_{0}},
\end{equation}
\begin{equation}
\gamma_{T}=m^{-2}T_{0}\frac{\partial m^{2}(T)}{\partial T_{0}},
\end{equation}
and
\begin{equation}
\Gamma^{(N)}=
\left.
\frac{\partial^{N}V_{\mbox{eff}}}{\partial\sigma_{0}^{N}}
\right|_{\sigma_{0}=0}.
\end{equation}
Using for the effective potential the expression (\ref{effective-potential}) 
we easily get
\begin{equation}
\beta_{T}=
\frac{1}{16\pi^{2}}
\left(
\frac{5}{12}\tilde{\lambda}^{2}+\tilde{g}^{4}
\right),
\label{beta-function}
\end{equation}
and
\begin{equation}
\gamma_{T}=
\frac{1}{12}
\left(
4\tilde{\lambda}-\tilde{g}^{2}
\right)
\frac{T^{2}}{m^{2}}+
\frac{\tilde{\lambda}}{4\pi^{2}}.
\label{gama-function}
\end{equation}

Equation (\ref{rg-equation}) can be solved by the method of characteristics.
One simply introduces a dilatation parameter $t$ and look for functions
$\lambda(t)$ (for fixed $g$) which satisfies
\begin{equation}
t\frac{d\lambda(t)}{dt}=\beta(\lambda(t)),
\end{equation}
with $\lambda(1)=\lambda$. If we now solve the above equation we get
\begin{equation}
\int_{\lambda}^{\lambda(t)}\frac{d\lambda^{'}}{\beta(\lambda^{'})}=ln{t}.
\label{running-coupling}
\end{equation}

To investigate the large $T$ limit we will have to study the behavior of the
effective coupling $\lambda(t)$ as $t$ goes to zero. Since $\beta(\lambda)$
is positive $\lambda(t)$ will decrease. Moreover, since the slope of
the $\beta$ function is also positive, the gaussian IR fixed point is
attractive. 

We can solve a similar equation to $g$ and defining 
$(\lambda^{*},g^{*})$ as the zero of the $\beta$ function
\begin{equation}
\beta(\lambda^{*},g^{*})=0,
\end{equation}
we see that, although the anomalous dimension for the $\sigma_{0}$ field
may change when we pass from the situation of total to the
situation of partial restoration of symmetry, the $\eta$
exponent defined as $\eta\equiv\eta(\lambda^{*},g^{*})$ 
remains unchanged $(\eta=0)$. 

One may ask whether renormalization group improves the critical
indices after dimensional reduction or not. The answer is no and
this should not be surprising since the form of the ultraviolet and
thermal divergences that we had to renormalize are of the same type
as for a theory in four dimensions, where everything is mean field.


\end{document}